\documentclass[aps,prl,twocolumn,superscriptaddress]{revtex4}
\usepackage{graphicx}
\usepackage{color}

\usepackage{amssymb}
\usepackage{amsmath}

\begin{document}


\title{High-harmonic probing of electronic coherence in dynamically aligned molecules}


\author{P. M. Kraus}
\affiliation{Laboratorium f\"ur Physikalische Chemie, ETH Z\"urich, Wolfgang-Pauli-Strasse 10, 8093 Z\"urich, Switzerland}
\author{S. B. Zhang}
\affiliation{Max Planck Institute for the Physics of Complex Systems, 01187 Dresden, Germany \\ and Center for Free-Electron Laser Science, DESY, 22607 Hamburg, Germany}
\author{A. Gijsbertsen}
\affiliation{Laboratorium f\"ur Physikalische Chemie, ETH Z\"urich,
Wolfgang-Pauli-Strasse 10, 8093 Z\"urich, Switzerland}
\author{R. R. Lucchese}
\affiliation{Department of Chemistry, Texas A\&M University, College Station, Texas 77843-3255, USA}
\author{N. Rohringer}
\affiliation{Max Planck Institute for the Physics of Complex Systems, 01187 Dresden, Germany \\ and Center for Free-Electron Laser Science, DESY, 22607 Hamburg, Germany}
\author{H. J. W\"orner}
\email[]{woerner@phys.chem.ethz.ch}
\homepage[]{www.atto.ethz.ch}
\affiliation{Laboratorium f\"ur Physikalische Chemie, ETH Z\"urich, Wolfgang-Pauli-Strasse 10, 8093 Z\"urich, Switzerland}


\date{\today}



\begin{abstract} We introduce and demonstrate a new approach to measuring coherent electron wave packets using high-harmonic spectroscopy. By preparing a molecule in a coherent superposition of electronic states, we show that electronic coherence opens previously unobserved high-harmonic-generation channels that connect distinct but coherently related electronic states. Performing the measurements in dynamically aligned nitric oxide (NO) molecules we observe the complex temporal evolution of the electronic coherence under coupling to nuclear motion. Choosing a weakly allowed transition to prepare the wave packet, we demonstrate an unprecedented sensitivity that arises from optical interference between coherent and incoherent pathways. This mechanism converts a 0.1 $\%$ excitation fraction into a $\sim$20 $\%$ signal modulation. \end{abstract}

\maketitle

Measuring the motion of valence-shell electrons in molecules is one of the central goals of modern ultrafast science. The last decade has witnessed very fundamental progress in this area with the development of attosecond streaking \cite{drescher02a,schultze10a} and interferometric techniques \cite{mauritsson08a} to time resolve electronic dynamics as well as transient absorption \cite{loh07a,goulielmakis10a} and strong-field ionization \cite{eckle08a,pfeiffer11a,woerner11a,fleischer11a} to probe electronic wave packets in atomic ions. All of these experiments have been performed on highly-excited states, in the continuum or in ionic species. Electronic dynamics involving the ground state and a low-lying electronically excited state of a neutral molecule have not been observed to date.

Here, we introduce a new all-optical technique that allows the measurement of an electronic wave packet in the valence shell of a neutral molecule for the first time. In our pump-probe experiment, an electronic wave packet is created through stimulated Raman scattering and probed by the generation of high-order harmonics (orders 9 to 23) of an infrared laser pulse. Figure 1A illustrates the concept of our measurement. High-harmonic emission from a coherent superposition of two electronic states can be described as a superposition of radiation produced in four channels illustrated by arrows in Fig.\ 1A. Ionization from and recombination to the same electronic state gives rise to two channels (blue arrows) that are independent of the electronic coherence. Ionization from one state and recombination to the other state gives rise to two additional channels (red arrows) that only contribute to an observable high-harmonic signal if the two states are coherently related. These channels are the key to probing electronic coherence and have not been observed previously. 

The two channels connecting the same initial and final states (blue arrows) emit radiation that is insensitive to the quantum phase of the initial state. The amplitude of the radiation generated in each of these channels is proportional to the {\bf population} in each state. These pathways have been exploited to time resolve photochemical dynamics \cite{woerner10b,woerner11c}. In contrast, the two cross-channels (red arrows) read out the relative quantum phases and encode their difference in the phase of the emitted radiation. The emission amplitude from these pathways is proportional to the product of the {\bf wave-function coefficients}, providing a very high sensitivity to weak excitations. 

We exploit this mechanism to reveal the sensitivity of high-harmonic spectroscopy to electronic coherence. Previous work showed that within a multi-channel approach allowing only single-particle hole excitations, the high-harmonic signal emitted from a molecule initially in the electronic ground state can be described as a superposition of the emission related to multiple channels connecting the same initial and final state through different continua (i.e. different states of the cation and associated photoelectron) \cite{mcfarland08a,smirnova09b,haessler10a,farrell11a}. Since different emission pathways can be realized in different emitters without changing the observed interference, just as in mixed gases \cite{kanai07a}, such measurements are insensitive to the electronic coherence between the electronic states of the cation. 


High-harmonic generation from coherently populated two-level systems has been intensively studied in theory (see e.g. \cite{millack93a,gauthey95a,sanpera96a,watson96a,niikura05a,milosevic06a}). Very recently, pump-probe schemes for coherent electronic wave packets in molecules have been studied numerically \cite{bredtmann11a,chelkowski12a}. In a pioneering experiment, incoherently prepared excited states have been shown to enhance high-harmonic emission but the effects of coherence could not be studied \cite{paul05a}. For simplicity, we assume here that the energy difference between the two electronic states $\Delta E$ is much smaller than their ionization energies and contained within the bandwidth of the laser pulse generating the high harmonics. The molecule is initially prepared in the electronic superposition state $\Psi=|c_1|e^{{\rm i}\phi_1}\psi_1+|c_2|e^{{\rm i}\phi_2}\psi_2$ where $\psi_1$ and $\psi_2$ are the normalized total wave functions of the two electronic states. The total emitted electric field at a given high-harmonic frequency can then be written as (see Fig.\ 1A and supplementary material, section I)
\begin{align} \label{fields}
E=&|c_1|^2 a_1 d_1+|c_2|^2 a_2 d_2 \nonumber \\
  &+|c_1c_2|e^{{\rm i}(\phi_1-\phi_2)}a_1 d_2+|c_1c_2|e^{{\rm i}(\phi_2-\phi_1)}a_2 d_1,
\end{align}
where $a_j$ stands for the complex spectral representation of the recombining photoelectron wave packet generated from level $j$ which includes the phase accumulation by laser-induced acceleration and $d_j$ stands for the complex photorecombination matrix elements. Since the wave-packet evolution during high-harmonic generation (i.e. one optical cycle) can be neglected under the present assumptions, no other frequencies in addition to the regular odd harmonics are generated. 


The experimental setup consists of an amplified femtosecond titanium:sapphire laser system and a vacuum chamber for generation and spectral characterization of high-order harmonic radiation. The output of the laser system (800 nm, 8 mJ, 25 fs, 1 kHz) is divided into pump and probe pulses separated by a controlled delay. The two beams are aligned parallel to each other with a vertical offset of 0.7 cm and are focused into a molecular beam generated by expansion of a 5 \% mixture of NO in helium through an Even-Lavie valve ($\o$ 150 $\mu$m) with a backing pressure of 9 bars. The pump pulses (60 fs) prepare the electronic and rotational dynamics of the NO molecules through impulsive stimulated Raman scattering (ISRS) while the probe pulses (30 fs) generate high-order harmonics with a cutoff at H23. The peak intensities of pump and probe pulses are estimated to (5$\pm1)\times 10^{13}$W/cm$^2$ and (1.5$\pm0.2)\times 10^{14}$W/cm$^2$ on target, respectively. We have not detected any even-harmonic emission from NO under the same experimental conditions that were used for the field-free orientation of OCS \cite{kraus12c}. The high harmonics generated by the probe beam propagate into an XUV spectrometer consisting of a 120 $\mu$m wide entrance slit, a concave aberration-corrected grating (Shimadzu, 30-002) and a microchannel-plate detector backed with a phosphor screen. The polarization of the probe pulse is kept unchanged (parallel to the grooves of the grating), whereas the polarization of the pump pulse is varied. A charge-coupled device camera records the spectral images and transfers them to a computer for analysis. 

The NO molecule has two low-lying electronic states $^2\Pi_{1/2}$ and $^2\Pi_{3/2}$ separated by $\sim$120 cm$^{-1}$. The eigenstates of the coupled rotational-electronic Hamiltonian are designated by F$_1$ ($|\Omega|=1/2$, dominated by the $^2\Pi_{1/2}$ state) and F$_2$ ($|\Omega|=3/2$, dominated by the $^2\Pi_{3/2}$ state), respectively, with $\Omega$ the projection quantum number of the total angular momentum on the internuclear axis. The molecules are initially prepared exclusively in the ground electronic (F$_1$) and vibrational state through supersonic expansion into vacuum with a rotational temperature of 15 K determined through comparison with simulations. An intense 60-fs laser pulse is used to transfer population from the F$_1$ to the F$_2$ state through ISRS. In NO, this transfer is only weakly allowed \cite{rasetti30a,lepard70a} such that 0.1-0.2 \% of the molecules are typically electronically excited (calculations are described in the supplementary material Section III). The weakness of the transition explains why none of the previous strong-field studies of NO has detected population in the F$_2$ state \cite{hasegawa06a,meijer07a,ghafur09a}. We note that the X~$^1\Sigma^+$ and the a~$^3\Sigma^+$ states of the cation are separated by 6.6~eV \cite{kimura81a}. We can thus safely neglect ionization to electronically excited states as well as laser-driven population transfer among the electronic states of the cation.

Figure 1B shows the integrated intensity of harmonic orders 9 to 23 (H9-H23) as a function of the time delay between the excitation and the probe pulses. We observe a highly regular quantum beat with a modulation depth of $\sim$ 30\% (defined as (max-min)/max). The measured period of 275$\pm$3 fs is in close agreement with the energy separation of the two electronic states. Since the population of the F$_2$ state is much smaller than that of the F$_1$ state ($|c_2| \ll |c_1| \approx 1$), the harmonic intensity is $I\approx\left|a_1d_1+c_2^*a_1d_2+c_2a_2d_1\right|^2$. Because the electronic structure of the F$_1$ and F$_2$ states only differs in the relative orientations of their orbital and spin angular momenta, the two states have virtually identical photorecombination transition moments ($d_2 \approx d_1 = d$). Since, in addition, the energy separation of the two states (0.015 eV or 120 cm$^{-1}$) is much smaller than the ionization energy (9.26 eV) the ionization rates are also very similar ($a_2\approx a_1 = a$). We thus obtain 
\begin{equation} \label{int}
I(t)\approx |ad|^2\left(1+2|c_2|\cos{(\Delta Et/\hbar)}\right)^2.
\end{equation}
The calculated population transfer of 0.17\% ($c_2=\sqrt{0.0017}$) thus gives rise to a modulation depth of 28.5 \%, in excellent agreement with the results shown in Fig.\ 1B.


We now show that our measurements further reveal the evolution in the orientation of the electronic wave packet. Figure 2 shows measurements recorded with parallel polarizations of the pump and probe pulses (red curves) or with perpendicular polarizations (blue curves). At early pump-probe delays (3-4 ps), we observe a simple sinusoidal modulation of the high-harmonic signal with the same phase and periodicity in all harmonic orders. Comparing the two polarization geometries, we find a $\pi$ phase shift and a smaller modulation depth in perpendicular polarizations ($\sim 7 \%$ in H15) than in parallel polarizations ($\sim 27 \%$ in H15). In the range of 4-6 ps, we observe the first half revival of the rotational dynamics. The pump pulse used to electronically excite the molecules also dynamically aligns them as we further discuss below.

The polarization dependence of both phase and contrast of the electronic quantum beat reveal the orientational evolution of the electronic wave packet. In the following, we discuss separately the electronic and rotational parts, because the total molecular wave functions of all levels of NO populated in the present experiment are well described by Hund's case (a) \cite{zare88a}, i.e. by products of electronic, rotational and vibrational wave functions. The wave functions of the $^2\Pi_{1/2}$ and $^2\Pi_{3/2}$ states are products of spatial and spin parts characterized by the projection quantum numbers of the electron orbital ($\Lambda$) and spin ($\Sigma$) angular momenta $|^2\Pi_{1/2},\pm\rangle=1/{\sqrt{2}}\left(|\pi^+\beta\rangle\pm|\pi^-\alpha\rangle\right)$ and $|^2\Pi_{3/2},\pm\rangle=1/{\sqrt{2}}\left(|\pi^+\alpha\rangle\pm|\pi^-\beta\rangle\right)$, where $\pm$ refers to the parity of the electronic level, $|\pi^{\pm}\rangle$ stands for $|\Lambda=\pm 1\rangle$ and $|\alpha\rangle/|\beta\rangle$ stands for $|\Sigma=\pm 1/2\rangle$. The time-dependent electron density $|\Psi^+(t)|^2$ of the coherent superposition of the $^2\Pi_{1/2}$ and $^2\Pi_{3/2}$ levels with positive parity is represented in Fig.\ 3. The electronic density corresponding to each of the two values of $\Sigma$ undergoes opposite rotations around the molecular axis. The coherent superposition state prepared in the experiment thus corresponds to two spin-density currents rotating around the molecular axis in opposite directions.

This representation can be used to analyze the sensitivity of the high-harmonic signals to the orientation of the electronic wave packet. Although high-harmonic generation is sensitive to the bound electronic wave function including phases \cite{itatani04a, haessler10a}, we find that the temporal evolution of the electronic density provides an intuitive qualitative explanation of the observed dynamics. A more quantitative description will be given in \cite{zhang13a}. Selection rules restrict efficient electronic excitation to molecules lying perpendicular to the pump polarization ($\Delta\Lambda=\pm 2$, $\Delta\Sigma=0$), while parallel molecules remain in the electronic ground state ($\Delta\Lambda=\Delta\Sigma=0$). 

We define the pump-pulse polarization as z-axis (Fig. 3) and first assume the molecule to lie parallel to the x-axis. The efficiency of high-harmonic generation is determined by both the strong-field ionization rate and the photorecombination dipole moments. Both quantities are suppressed along nodal planes of electronic orbitals and maximized along the directions of highest amplitude \cite{muthbohm00a,kanai05a,itatani05a}. In parallel polarizations of pump and probe, the maximal emission is thus observed when the electronic density is aligned along the direction of the probe polarization. At the same time, a minimum will be observed in perpendicular polarizations because the probe polarization lies in the nodal plane of the electronic density distribution. This explains the $\pi$ phase shift between the electronic quantum beat observed for different polarizations in Fig.\ 2. 

We now turn to the different modulation depths observed in Fig.\ 2. The electronic dynamics is expected to modulate the high-harmonic signal most strongly when the molecule lies perpendicular to the probe-pulse polarization, as illustrated in Fig.\ 3, but no modulation is expected when the molecule lies parallel to it. In the experiment featuring parallel pump-probe polarizations, the molecules predominantly lie perpendicular to the probe pulse resulting in a deep signal modulation. In perpendicular pump-probe polarizations, all angles between the excited molecules and the probe polarization are equally likely, which explains the smaller modulation depth. These simple considerations thus explain both the observed phase shift and the different modulation depths. 



We have chosen a molecule for our study of valence-shell electronic coherences to also explore the manifestations of nuclear-electronic coupling in high-harmonic spectroscopy. Figure 4A shows the intensities measured in H9 and H15. In addition to the electronic quantum beat visible in all harmonic orders, three prominent rotational revivals are also visible, most clearly in H9. Whereas H9 is most sensitive to the rotational motion, H15 and higher harmonics are mainly sensitive to the electronic motion. This observation is rationalized by the angular variation of photorecombination dipole moments obtained from quantum scattering calculations following Ref. \cite{stratmann96a} and displayed in the inset of Fig.\ 4D. While H9 displays a pronounced amplitude dependence on the alignment angle, the angular dependence of H15 is much weaker, explaining the observed reduction in the sensitivity of H15 and higher harmonic orders to the rotational dynamics.

The sinusoidal oscillations observed in the first 4 ps become less regular and less pronounced with increasing pump-probe delays, reaching a minimal modulation depth around 15.5 ps followed by an increase of the modulation depth towards a delay of 18 ps (see also supplementary material, section IV). Figure 4B shows the Fourier-transform power spectra of Fig.\ 4A. The single peak centred at 120$\pm$0.5 cm$^{-1}$ corresponds to the electronic coherence whereas the additional peaks observed between 20 and 60 cm$^{-1}$ correspond to rotational coherences in the F$_1$ state associated with a change of 2 units in the total angular momentum ($\Delta J=\pm 2$). A more detailed assignment and a level diagram of NO are given in the supplementary material (Figs. S2, S3). The observation of a single intense peak at 120 cm$^{-1}$ shows that the dominant modulation is caused by purely electronic transitions without change of the rotational state ($\Delta J=0$) in the high-harmonic generation process as illustrated in Fig.\ 1A. The much weaker peaks between 130 and 150 cm$^{-1}$ correspond to mixed electronic-rotational transitions with $\Delta J=+1$ that we further discuss below.

We support our observations and assignments with calculations solving the time-dependent Schr\"odinger equation for the coupled rotational and electronic motions under the experimental conditions (details are given in the supplementary information, Section III, and in a forthcoming publication \cite{zhang13a}). These calculations provide the degree of alignment of the molecule ($\langle\cos^2\theta\rangle$, Fig.\ 4C) which enables the identification of the rotational revivals observed in the experiment. We further calculate the modulation of the high-harmonic signal arising from the electronic motion using a reduced density matrix formalism. Briefly, we construct an electronic density matrix in the basis of the two electronic states by tracing over the rotational quantum states and calculate the high-harmonic yield using the reduced density-matrix equivalent of Eq. (2), i.e. $I(t)\propto[1+2\Re(\rho^R_{12}(t))]^2$, where $\rho^R_{12}$ is the off-diagonal element of the reduced density matrix. This calculation predicts a damped electronic quantum beat that is shown in the lower panel of Fig.\ 4C. Figure 4D shows the Fourier-transform power spectra of Fig.\ 4C. The top panel shows dominant $\Delta J=\pm 2$ rotational coherences and the bottom panel shows a dominant electronic coherence. The weaker mixed coherences observed in the experiment are not predicted by our reduced density-matrix formalism which confirms their assignment to high-harmonic generation channels that connect levels differing in both electronic and rotational quantum numbers. Using a filtered back-transformation method, we show in the supplementary material (section IV and Fig.\ S1) that these mixed rotational-electronic transitions explain the complex dephasing and rephasing of the electronic quantum beat described above.


In conclusion, we have introduced a general new approach to probing quantum coherences and have used it to demonstrate the pronounced sensitivity of high-harmonic spectroscopy to electronic coherence. The observed signal modulations are explained by high-harmonic generation pathways that connect two coherently related electronic states and encode the difference of their quantum phases into the phase of the emitted radiation. The temporal evolution of the relative phase of the two levels is thus detected through interference between these coherent pathways and incoherent pathways connecting the same initial and final states. This new physical mechanism enhances extremely weak coherences to a readily observable level. By studying a molecule, we have shown that the electronic quantum beat evolves under the influence of mixed rotational-electronic coherences, resulting in dephasing and partial rephasing as a function of time. Although demonstrated on the femtosecond time scale and a specific molecule, our method is well suited to explore femtosecond and attosecond dynamics \cite{niikura05a,bredtmann11a,chelkowski12a} of any molecule featuring electronically excited states within the bandwidth of the pump pulse which can be made extremely broad \cite{wirth11a}. Stimulated Raman scattering may indeed become a useful approach to preparing coherent wave packets, giving access to  very broad spectral ranges, spanning not only rotational and vibrational levels but also several electronic states as demonstrated in the present study. Such experiments would be nearly ideal probes of molecular dynamics, revealing how electronic, vibrational and rotational dynamics unfold in the time domain while the spectral resolution can be arbitrarily improved through an extension of the measured range of pump-probe delays.

\begin{acknowledgments}
We gratefully acknowledge funding from the Swiss National Science Foundation (PP00P2\_128274) and ETH Z\"urich (ETH-33 10-3). We thank David Villeneuve, Kresimir Rupnik and Aaron von Conta for discussions.
\end{acknowledgments}


\begin{figure*}
\includegraphics[width=0.7\textwidth]{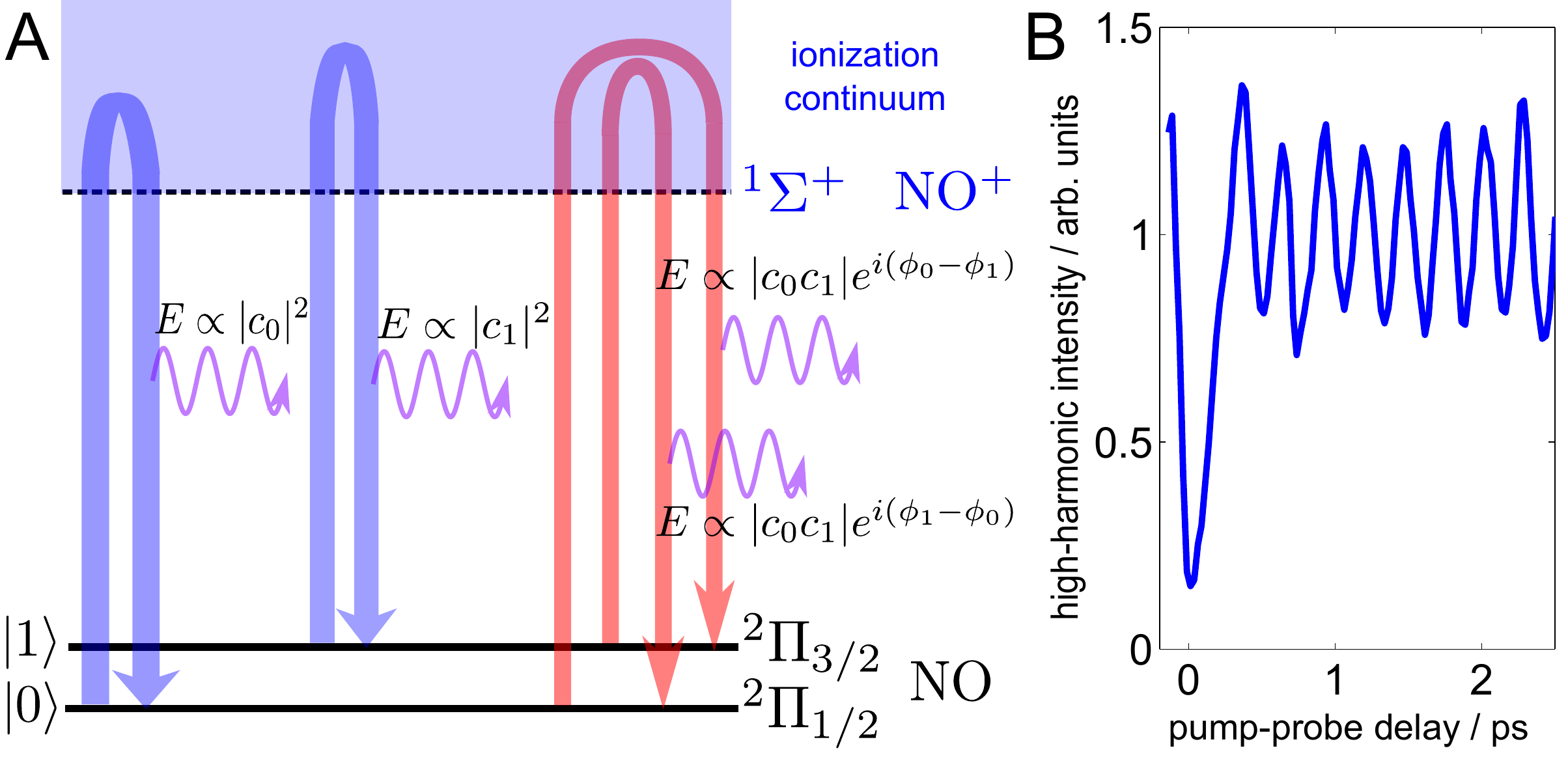}
\caption{A) High-harmonic generation form a coherent superposition of two electronic states $\Psi=c_1\psi_1+c_2\psi_2$ can be rationalized as the sum of two incoherent (blue arrows) and two coherent (red arrows) pathways. B) High-harmonic intensity integrated over the harmonic orders 9-23 emitted from a coherent superposition of the $^2\Pi_{1/2}$ and $^2\Pi_{3/2}$ states of nitric oxide as a function of the pump-probe delay (parallel pol.).}
\end{figure*}

\begin{figure*}
\includegraphics[width=0.7\textwidth]{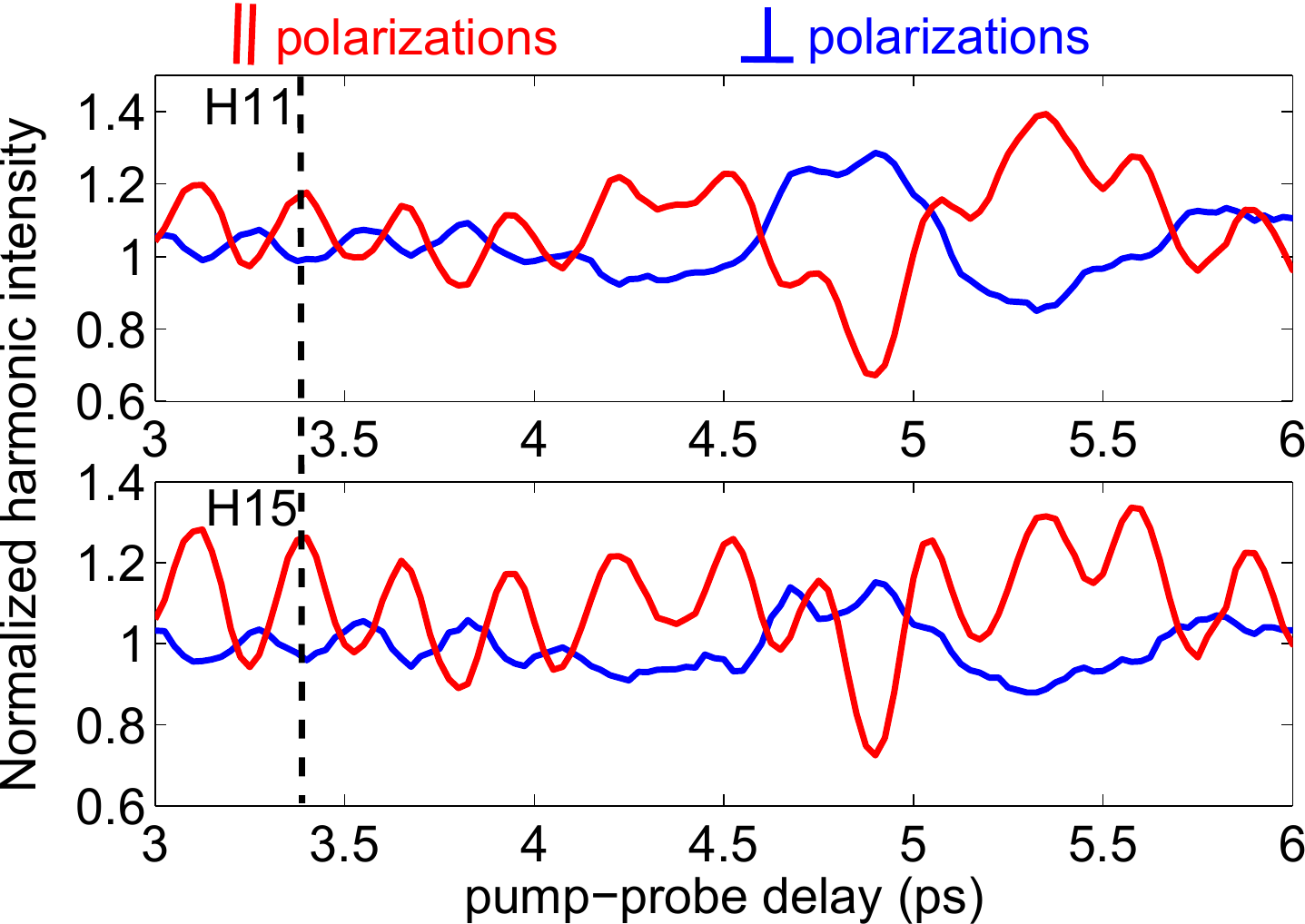}
\caption{Measured high-harmonic intensities for parallel and crossed polarizations of pump and probe pulses. The region from 3-4 ps is dominated by the electronic quantum beat whereas the region from 4-6 ps reveals the first half-revival of the molecular alignment.}
\end{figure*}

\begin{figure*}
\includegraphics[width=0.6\textwidth]{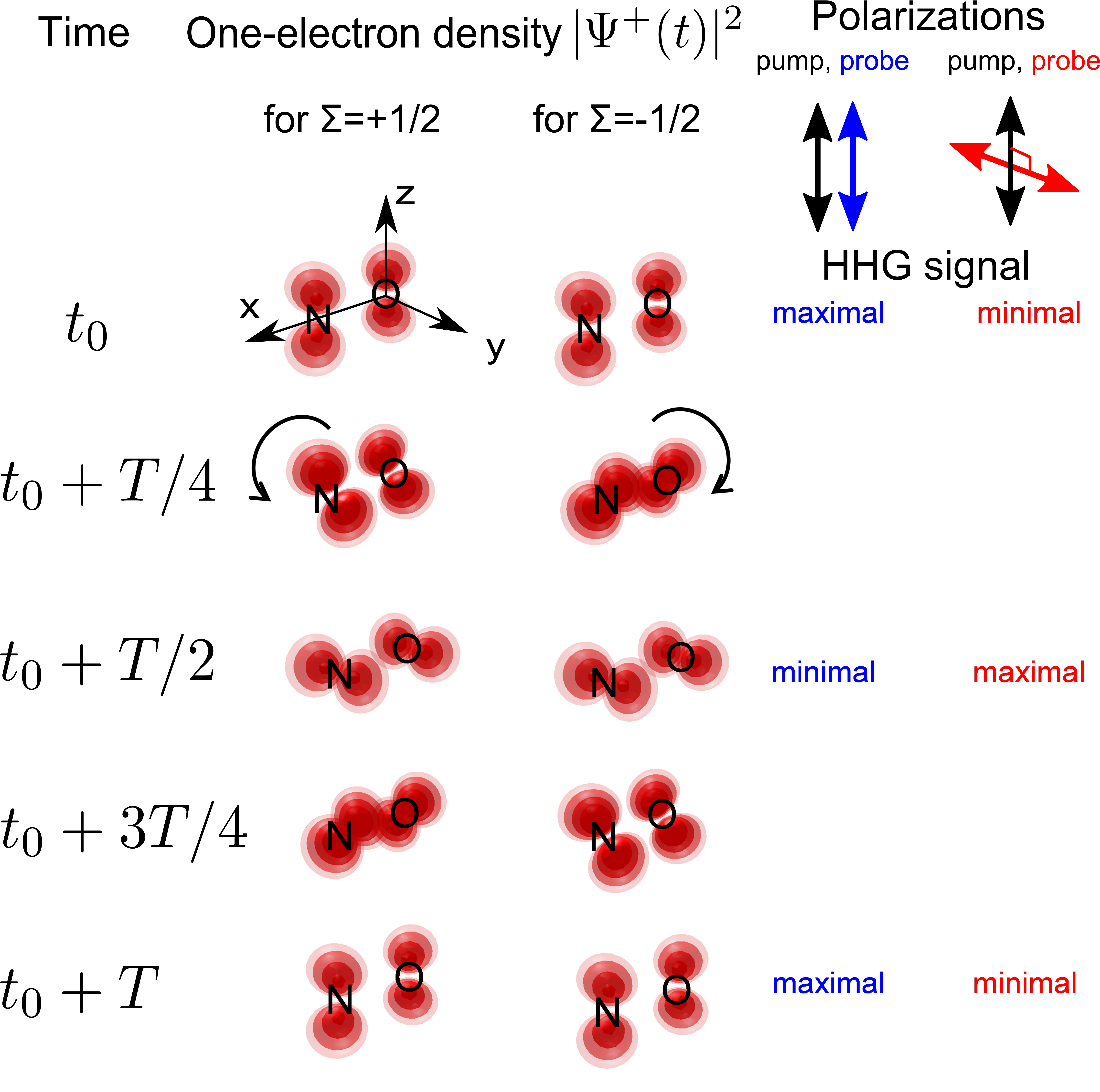}
\caption{Temporal evolution of the electronic density corresponding to the coherent superposition state $|\Psi^+(t)\rangle=|^2\Pi_{1/2},+\rangle+|^2\Pi_{3/2},+\rangle e^{-{\rm i}\Delta E t/\hbar}$ prepared by a pump pulse linearly polarized along z and expected efficiency of high-harmonic generation for different linear probe polarizations (z or y). The electronic densities corresponding to $\Sigma=\pm 1/2$ are shown separately. $T$ represents the period of the electronic wave packet.}
\end{figure*}

\begin{figure*}
\includegraphics[width=0.8\textwidth]{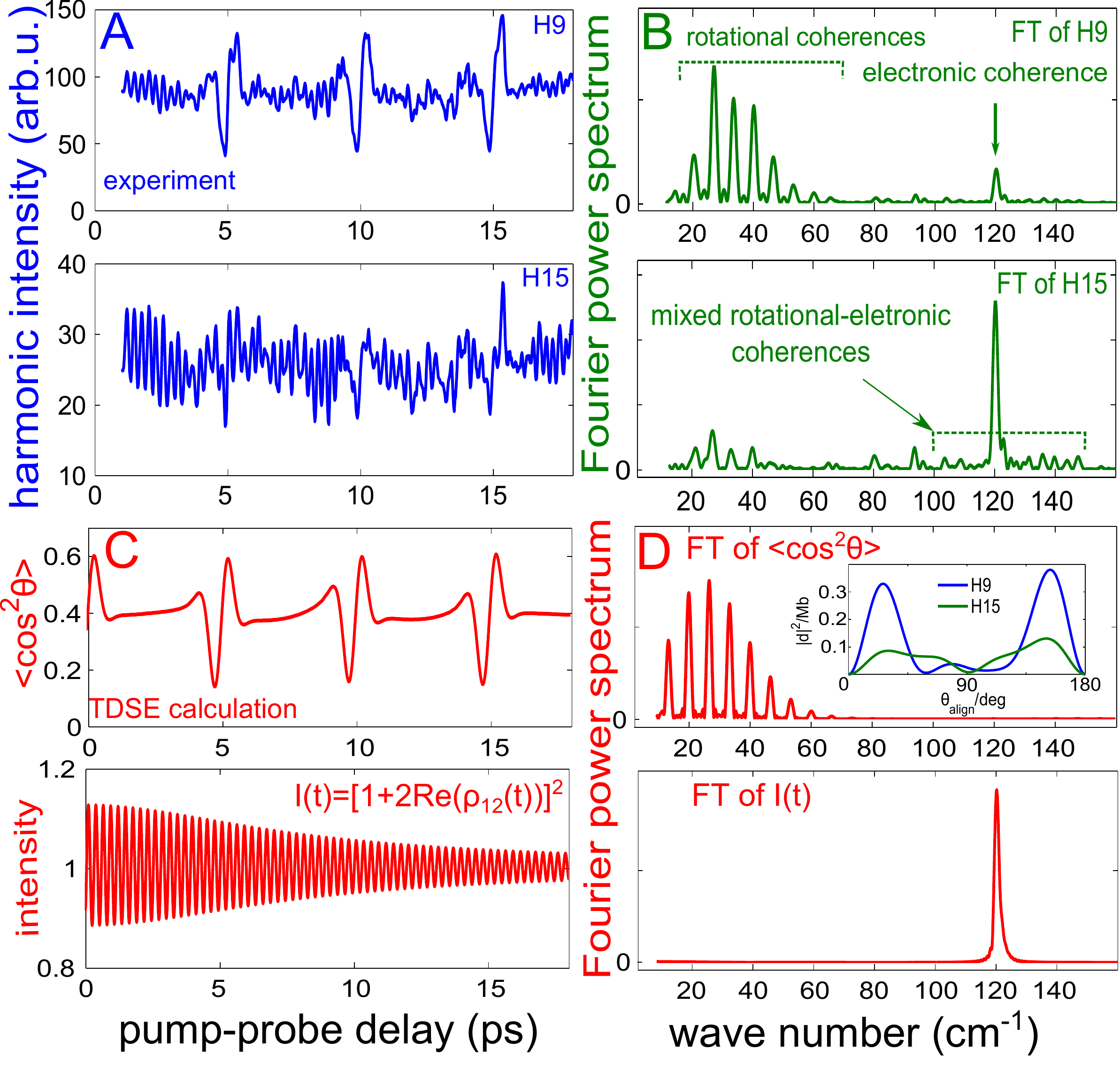}
\caption{Identification of electronic, rotational and mixed coherences. A) Measured intensities under parallel polarizations. B) Fourier-transform power spectra of traces in A. C) Degree of axis alignment $\langle\cos^2\theta\rangle$ and intensity modulation resulting from the electronic motion, obtained by solving the time-dependent Schr\"odinger equation D) Fourier-transform power spectra of the calculated traces in C. The inset shows calculated molecular-frame photorecombination dipole moments for H9 and H15.} 
\end{figure*}

\end{document}